\documentclass[letter]{aa} 

\usepackage{pgf}
\usepackage{graphicx}
\usepackage{txfonts}
\usepackage{natbib,twoopt}
\usepackage[breaklinks=true]{hyperref} 
\bibpunct{(}{)}{;}{a}{}{,} 

\usepackage{graphicx}
\usepackage{txfonts}
\begin{document}

   \title{The January 2015 outburst of a red nova in M31}

   \author{
Alexander Kurtenkov\inst{1,8}
\and
Peter Pessev\inst{2,7}
\and
Toma Tomov\inst{3}
\and
Elena A. Barsukova\inst{4}
\and
Sergei Fabrika\inst{4,15}
\and
Kriszti\'{a}n Vida\inst{5}
\and
Kamil Hornoch\inst{6}
\and
Evgeni P. Ovcharov\inst{1}
\and
Vitaly P. Goranskij\inst{9}
\and
Azamat F. Valeev\inst{4,15}
\and
L\'{a}szl\'{o} Moln\'{a}r\inst{5}
\and
Kriszti\'{a}n S\'{a}rneczky\inst{5}
\and
Andon Kostov\inst{8}
\and
Petko Nedialkov\inst{1}
\and
Stefano Valenti\inst{10}
\and
Stefan Geier\inst{2,7}
\and
Klaas Wiersema\inst{11}
\and
Martin Henze\inst{12,16}
\and
Allen W. Shafter\inst{13}
\and
Rosa Victoria Mu\~{n}oz Dimitrova\inst{8}
\and
Vasil N. Popov\inst{8}
\and
Maximilian Stritzinger\inst{14}
}

\institute{
Department of Astronomy, University of Sofia, 5 James Bourchier Blvd., 1164 Sofia, Bulgaria, \href{mailto:al.kurtenkov@gmail.com}{\url{al.kurtenkov[at]gmail.com}}
\and
Instituto de Astrof\'{i}sica de Canarias, 38200 La Laguna, Tenerife, Spain, \href{mailto:peter.pessev@gtc.iac.es}{\url{peter.pessev[at]gtc.iac.es}}
\and 
Centre for Astronomy, Faculty of Physics, Astronomy and Informatics, Nicolaus Copernicus University, Grudziadzka 5, 87-100 Torun, Poland, \href{mailto:tomov@umk.pl}{\url{tomov[at]umk.pl}}
\and 
Special Astrophysical Observatory, Nizhnij Arkhyz, Russia
\and
Konkoly Observatory, Hungarian Academy of Sciences, H-1121 Budapest, Konkoly Thege Mikl\'os \'ut 15-17, Hungary 
\and
Astronomical Institute, Academy of Sciences, CZ-251 65 Ond\v{r}ejov, Czech Republic
\and
Departamento de Astrof\'{i}sica, Universidad de La Laguna, E-38206 La Laguna, Tenerife, Spain
\and
Institute of Astronomy and NAO, Bulgarian Academy of Sciences, 
72 Tsarigradsko Chaussee Blvd., 1784 Sofia, Bulgaria
\and
Sternberg Astronomical Institute, M.V. Lomonosov Moscow State University, Universitetskii pr. 13, 119992 Moscow, Russia
\and
Las Cumbres Observatory Global Telescope Network, 6740 Cortona Dr., Suite 102, Goleta, CA 93117, USA
\and
Department of Physics and Astronomy, University of Leicester, Leicester, LE1 7RH, UK
\and
European Space Astronomy Centre, P.O. Box 78, 28692 Villanueva de la Ca\~{n}ada, Madrid, Spain
\and
Department of Astronomy, San Diego State University, San Diego, CA 92182, USA
\and
Department of Physics and Astronomy, Aarhus University, Ny Munkegade 120, DK-8000 Aarhus C, Denmark
\and
Kazan Federal University, Kazan, 420008, Russia
\and
Institut de Ci\`encies de l'Espai (CSIC-IEEC), Campus UAB, C/Can Magrans s/n, E-08193 Cerdanyola del Valles, Spain
}

   \date{Received May 20, 2015; accepted May 28, 2015}

 
  \abstract
   {M31N 2015-01a (or M31LRN 2015) is a red nova that erupted in January 2015 -- the first event of this kind observed in M31 since 1988. Very few similar events have been confirmed as of 2015. Most of them are considered to be products of stellar mergers. }
   {Results of an extensive optical monitoring of the transient in the period January-March 2015 are presented.}
   {Eight optical telescopes were used  for imaging. Spectra were obtained on BTA, GTC and the Rozhen 2m telescope.}
   {We present a highly accurate 70 d lightcurve and astrometry with a 0.05\arcsec uncertainty. The color indices reached a minimum 2-3 d before peak brightness and rapidly increased afterwards. The spectral type changed from F5I to F0I in 6 d before the maximum  and then to K3I in the next 30 d.  The  luminosity of the transient was estimated to $8.7^{+3.3}_{-2.2}\times10^{5}L_{\odot}$ during the optical maximum.}
   {Both the photometric and the spectroscopic results confirm that the object is a red nova, similar to V838 Monocerotis.}


   \maketitle
%

\section{Introduction}

Red novae (or red transients) are a rare class of optical transients, reaching a peak luminosity equal to or higher than the brightest classical novae, but still lower than supernovae. Many of them are best explained by binary star mergers \citep{2006A&A...451..223T} and therefore are often referred to as stellar mergers or "mergebursts" \citep{2011A&A...528A.114T,2014MNRAS.443.1319K}. Common observational properties during the first months of the outburst are the initial slow decline and a spectrum changing towards a late-type supergiant phase with colors shifting to the red.

Modern observational history of these events goes back to 1988 when the transient M31-RV was detected by \citet{1989ApJ...341L..51R} as an M0 supergiant variable with peak absolute $M_{bol}\sim-10$. A similar outburst in the Milky Way was observed in 1994 -- V4332 Sgr changed its spectrum from K3 to M8-9 during a 3-month period \citep{1999AJ....118.1034M}. The eruption of V838 Mon in February 2002 \citep{2002A&A...389L..51M} reached an apparent magnitude of V=6.7\,mag and was studied quite closely. The lightcurve showed three maxima and a consistent change of colors redwards as the spectrum reached an L supergiant phase at the end of the year \citep{2003MNRAS.343.1054E}. The resulting circumstellar light echo was used to obtain a distance estimate of 6.1$\pm$0.6\,kpc, corresponding to a peak absolute magnitude $M_V=-9.8$ \citep{2003Natur.422..405B,2008AJ....135..605S}. A bright (peak $M_R\sim-12$) event, M85 OT2006-1, was reported by \citet{2007Natur.447..458K}. Like other red novae, it showed a narrow H$\alpha$ emission, a temperature decrease and an infrared excess a few months after the eruption \citep{2007ApJ...659.1536R}. 

V1309 Sco is another Milky Way red nova, discovered in 2008. \citet{2011A&A...528A.114T} have shown that the progenitor is a contact binary with a 1.4-day period, thereby adding a significant argument in favour of the stellar merger model for the cause of these outbursts. 

It has been suggested that the first catalogued nova, CK Vul was also a "red transient" \citep{2015arXiv150306570K}. A possible red nova in M101 was reported by \citet{2015ATel.7206....1G}. 

Here, we present a new addition to this class of objects, a red nova that erupted in M31 in Jan 2015. M31N 2015-01a was discovered by the MASTER-Kislovodsk auto-detection system on 2015 Jan 13 at $\sim$19.0\,mag, unfiltered \citep{2015ATel.6911....1S}. It was incorrectly identified as a classical nova by \citet{2015ATel.6941....1K} based on a strong H$\alpha$ emission on the Jan 16 spectrum, also reported by \citet{2015ATel.6952....1H}. On Jan 15 the object showed the spectral characteristics of an F5 supergiant \citep{2015ATel.6985....1F}. The transient became very bright, reaching a peak magnitude of R$\sim$15.1\,mag circa Jan 22. It then proceeded to fade slowly and consistently turned redder. At the end of February the colors had changed considerably with the spectrum resembling that of a K supergiant. Subsequently, we announced that the transient is a red nova \citep{2015ATel.7150....1K}. Our supposition that the overall spectral evolution of the object is very similar to the observed in \object{V838~Mon} was confirmed by \citet[][Fig.~2]{2015arXiv150407747W}. The possible progenitor system has been discussed by \citet{2015ATel.7173....1D} and \citet{2015arXiv150407747W}.


\section{Observations and data reduction}

We performed imaging of M31N 2015-01a in Johnson-Cousins $BVRI$ filters (Table\,\ref{telescopes}). The science frames were divided by flat fields after dark/bias subtraction. 

\begin{table}
\begin{center}
\caption[Telescopes used for the monitoring of M31N 2015-01a. The last column contains the designations used in the online Table\,\ref{table_phot}.]{Telescopes used for the monitoring of M31N 2015-01a. The last column contains the designations used in the online Table\,\ref{table_phot}.}\label{telescopes}
\small
\begin{tabular}{l@{}l@{ }c@{ }c@{ }c@{ }c@{ }c@{ }c@{ }c@{}}
\hline\hline
Telescope  & Observatory  & Country  & Image scale \, &   Des. \\
           &        	  &	     & [arcsec/px]   &   	\\
\hline
6 m BTA  	 \, & SAO RAS			 \, & Russia	 \, & 0.36  &  BTA  \\
2.5 m NOT  	 \, & ORM		 	 \, & Spain      \, & 0.19  &  NOT  \\
2 m RCC  	 \, & Rozhen NAO		 \, & Bulgaria   \, & 0.74  &  R2M  \\
1 m LCOGT 	 \, & McDonald   \, & USA	 \, & 0.47 &  LCOGT  \\
65 cm  \, & Ond\v{r}ejov   \, & Czech Rep. \, & 1.05  &  ONDR  \\
60/90 cm Schmidt \, & Konkoly    \, & Hungary	 \, & 1.03  &  KSCH  \\
50/70 cm Schmidt \, & Rozhen NAO \, & Bulgaria	 \, & 1.08  &  RSCH  \\
50 cm Cassegrain \, & Uni Leicester \, & UK	 \, & 0.89  &  UL50  \\
\hline        
\end{tabular} 
\end{center} 
\end{table}   

Images of vastly different depths and FOVs were obtained. In order to do all the photometry in a similar manner, we used four bright reference stars, within 1.6\arcmin\ from the transient (Table\,\ref{refstars}). That way systematic effects over large FOVs are  not affecting  the results. This also allowed us to minimize the instrumental errors of the references. Three out of those four stars were used for each frame. \citet{2006AJ....131.2478M} found systematic differences of the order of 0.1\,mag between their magnitudes and the ones by \citet{1992A&AS...96..379M}, so we used the Massey catalog to recalibrate the $BVRI$ magnitudes of the four references. Two images per filter and between 18 and 30 stars per image were used for the recalibration. The peak signals of the four reference stars on those images are within the linear range of the detector. The results are presented in Table\,\ref{refstars}.

\begin{table}
\begin{center}
\caption[List of reference stars. The  magnitudes are calibrated by  the \citet{2006AJ....131.2478M} catalog using images from the 6m BTA, the Rozhen 2m and the Konkoly Schmidt telescopes.]{List of reference stars. The  magnitudes are calibrated by  the \citet{2006AJ....131.2478M} catalog using images from the 6m BTA, the Rozhen 2m and the Konkoly Schmidt telescopes.}\label{refstars}
\small
\begin{tabular}{l@{}l@{ }c@{ }c@{ }c@{ }c@{ }c@{ }c@{ }c@{}}
\hline\hline
 $\alpha$ (J2000) \,& $\delta$ (J2000) \,& B \,& V	\,& R \,& I 		     \\
\hline
  00:42:08.9 \,& +40:55:33 \,& 14.521 \,& 13.739 \,& 13.312 \,& 12.886	  \\
  00:42:12.6 \,& +40:54:38 \,& 14.792 \,& 14.267 \,& 13.996 \,& 13.699	  \\
  00:42:12.2 \,& +40:55:49 \,& 16.030 \,& 15.287 \,& 14.883 \,& 14.495	  \\
  00:42:05.3 \,& +40:53:38 \,& 16.876 \,& 15.915 \,& 15.381 \,& 14.822	 \\
\hline
rms of calibration:  \,&  \,& 0.015 \,& 0.007 \,& 0.008 \,& 0.015 \\
\hline        
\end{tabular} 
\end{center}  
\end{table}  

Aperture photometry was done for all individual frames. The aperture radii we used were in the range of 1--1.5 FWHM of the Gaussian profiles of point sources. The results are presented as online material in Table\,\ref{table_phot}. Only the mean JD and median magnitude are given where several consecutive images are available. 

\begin{figure}
\centering
\includegraphics[width=8.9cm]{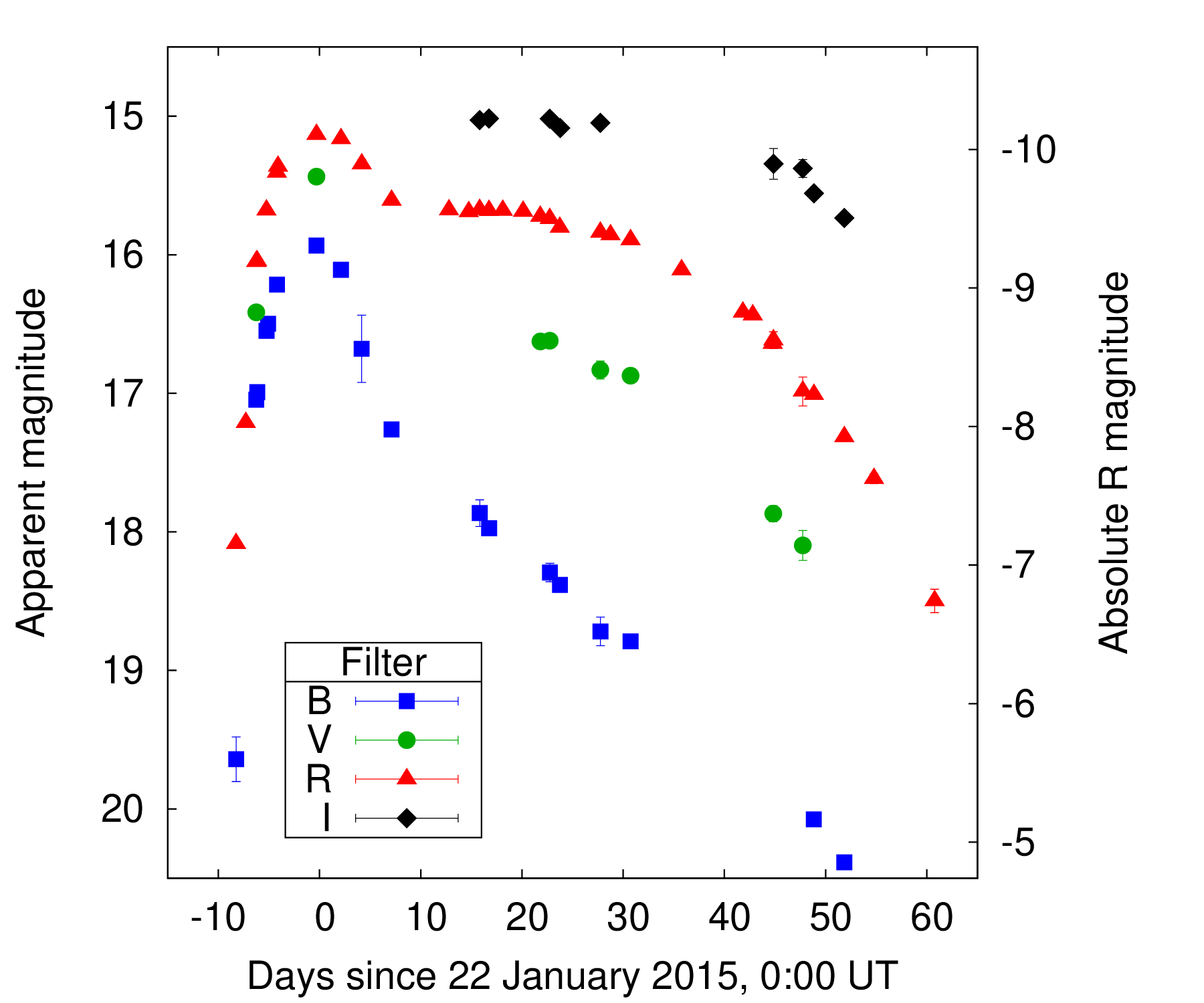}
\caption{Lightcurve of M31N 2015-01a in Johnson-Cousins $BVRI$ filters. The zero-point moment roughly coincides with the $R$-band peak. The absolute magnitude scale (right) applies to the $R$-band datapoints only.}
\label{lightcurve}
\end{figure}

Spectra  of \object{M31N 2015-01a} were obtained with three telescopes -- the 6 m BTA at SAO RAS, the 10.4 m GTC at ORM and the 2 m RCC telescope at Rozhen NAO (Table\,\ref{table_spec}). All the data reduction and calibrations were carried out with standard \textsc{midas}\footnote{See \url{http://www.eso.org/sci/software/esomidas/}} (BTA) and \textsc{iraf}\footnote{\textsc{IRAF} is distributed by the National Optical Astronomy Observatory, which is operated by the Association of Universities for Research in Astronomy (AURA) under a cooperative agreement with the National Science Foundation. } (2m RCC and GTC) procedures. The BTA and the GTC spectra were calibrated in relative fluxes using spectrophotometric standards \object{G191-B2B}, \object{Hiltner 600} and \object{Hz 2} \citep{1990AJ.....99.1621O,1992PASP..104..533H,1994PASP..106..566H}.

\begin{table}
\begin{center}
\caption[Log of spectral observations.]{Log of spectral observations. The instruments are described by \citet{2000KFNTS...3...13J}, \citet{2005AstL...31..194A} and \citet{1998Ap&SS.263..369C}.}\label{table_spec}
\small
\begin{tabular}{l@{}l@{ }c@{ }c@{ }c@{ }}
\hline\hline
Date  & Telescope/Instrument  & R  & Range [\AA]\, & Exp. time [s] \\
\hline
Jan 15  	 \, & BTA/SCORPIO		 \, & 1000	\, & 4050-5850 \, & 1500 \\
Jan 16  	 \, & 2mRCC/FoReRo2		 \, & 400      	\, & 5500-7500 \, & 4$\times$900 \\
Jan 17  	 \, & 2mRCC/FoReRo2		 \, & 400   	\, & 5500-7500 \, & 5$\times$900 \\
Jan 21  	 \, & BTA/SCORPIO		 \, & 1000	\, & 4050-5850 \, & 1800 \\
Feb 21  	 \, & BTA/SCORPIO		 \, & 1000	\, & 3720-5530 \, & 2700 \\
Feb 22  	 \, & BTA/SCORPIO		 \, & 1000	\, & 5750-7500 \, & 1800 \\
Feb 24  	 \, & GTC/OSIRIS		 \, & 2500	\, & 4430-9090 \, & 6$\times$180 \\
\hline        
\end{tabular} 
\end{center}  
\end{table}


\section{Results and discussion}

\subsection{Astrometry}

Astrometric calibrations were made using both the PPMXL \citep{2010AJ....139.2440R} and the LGGS \citep{2006AJ....131.2478M} catalogs. Astrometric solutions were obtained for 5 images from the 10.4 m GTC, 6 m BTA and 1 m LCOGT telescopes with an atmospheric seeing in the range of $1.3\arcsec-1.7\arcsec$. Separate calibrations were made using 20 relatively bright stars from the PPMXL catalog and 16 fainter stars close to the transient from LGGS. From PPMXL we derived $\alpha$,\,$\delta$\,(J2000) = 00\textsuperscript{h}42\textsuperscript{m}08\fs053, +40\degr55\arcmin01\farcs27 with uncertainties of $\Delta_\alpha\sim0.2\arcsec$, $\Delta_\delta\sim0.1\arcsec$. The LGGS calibrations yielded $\alpha$,\,$\delta$\,(J2000) = 00\textsuperscript{h}42\textsuperscript{m}08\fs065, +40\degr55\arcmin01\farcs33 with $\Delta_\alpha\sim\Delta_\delta\sim0.05\arcsec$.

\subsection{Photometry}

The light curve of \object{M31N 2015-01a} (Fig.~\ref{lightcurve}) shows a distinct similarity to the V1309 Sco lightcurve during the first month of the outburst \citep{2010A&A...516A.108M}. The Jan 21 datapoints are obtained shortly after the $B$-band and before the $R$-band maxima, so we assume that they coincide with the $V$-band maximum at 15.43\,mag. The initial fading has been quite slow. The decline times by 2 mag after the maximum $t_2$ were $\sim17\,d$, $\sim40\,d$ and $\sim50\,d$ in $B$, $V$ and $R$-bands respectively. The colors have been consistently shifting to the red, e.g. $B-R$ increased by $2.1$\,mag in the 31 days after the maximum. The $R$ magnitude reached a plateau 0.6\,mag below the maximum in $\sim$10\,d, but no rebrightening followed, as was the case with V838 Mon.   

\begin{figure}
\centering
\includegraphics[width=8.2cm]{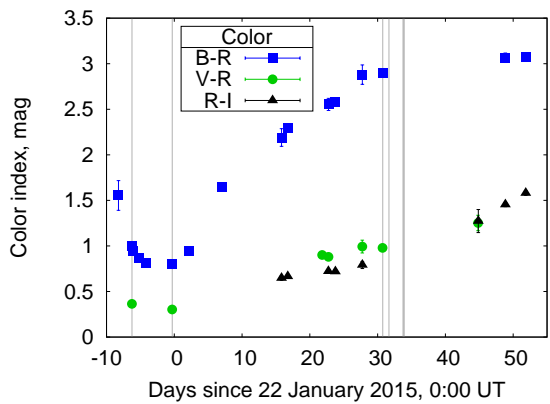}
\caption{The change of the observed color indices. Vertical lines represent the times the BTA (thin) and GTC (thick) spectra were taken at.}
\label{colors}
\end{figure}

Figure\,\ref{colors} shows the change of colors. The color temperature initially increased and reached a peak 2--3 d before the $R$-band maximum. Afterwards it has consistently decreased as the $B-R$ index increased by $\sim$2 mag in one month.

\subsection{Spectroscopy}

The  aim of the FoReRo2 spectra of \object{M31N 2015-01a} obtained 5 and 4 days before the maximum was to confirm that the object is a nova. The only features in these spectra are the emission lines of H$\alpha$ with equivalent widths $19\pm1$\,\AA\ and $17\pm1$\,\AA\ on Jan 16 and 17 respectively and a \ion{Na}{I} doublet absorption. After a careful re-calibration of the wavelength scales \citep[see][]{2015ATel.6941....1K} the measured radial velocities of the H$\alpha$ emission peaks are $-365$\,km\,s$^\mathrm{-1}$ and $-350$\,km\,s$^\mathrm{-1}$. In good agreement with the velocity of $\sim-370\pm50$\,km\,s$^\mathrm{-1}$ estimated by \citet{2015ATel.6985....1F} for the part of M31 where the object is located. 

\begin{figure}
\centering
\includegraphics[width=8.9cm]{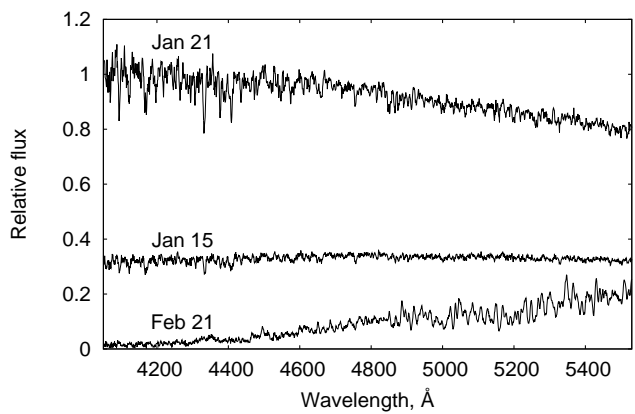}
\caption{BTA/SCORPIO spectra obtained with the VPHG1200B and G grisms. The Jan 21 spectrum is close to maximum brightness and arbitrary shifted to unity. The relative flux and shape of the Jan 15 and Feb 21 spectra reflect the evolution of the luminosity and the colors.} 
\label{btaspec}
\end{figure}

The first SCORPIO spectrum was obtained 6 days before the maximum brightness. The spectral region covered (4050-5850\,\AA) is dominated by absorption lines of ionized and neutral elements like \ion{Fe}{II}, \ion{Ti}{II}, \ion{Cr}{II}, \ion{Mg}{II} and \ion{Fe}{I}. H$\gamma$ and H$\delta$ are visible in absorption, while H$\beta$ absorption is partially filled by an emission component. A comparison of the absorption line spectrum with the \citet{1984ApJS...56..257J} library of stellar spectra shows very close similarity with a F5I spectrum. The next SCORPIO spectrum obtained on Jan 21 more or less coincides with the brightness maximum. The absorption line spectrum remained practically the same. However, the continuum is much bluer (Fig.~\ref{btaspec}) showing a better resemblance with a F0I spectrum.  We cross-correlated these two spectra with a F5I template and found a radial velocity for the pre-maximum spectrum $-462\pm38$\,km\,s$^\mathrm{-1}$ and $-534\pm14$\,km\,s$^\mathrm{-1}$ for the spectrum around brightness maximum (measured using the \textsc{iraf fxcor} package). The cross-correlation of the \object{M31N 2015-01a} spectra itself, confirms a velocity difference of $64\pm13$\,km\,s$^\mathrm{-1}$. Taking into consideration the velocity $-370$\,km\,s$^\mathrm{-1}$, connected with M31, we found that the heliocentric velocity of the ejected matter on Jan 15 was $\sim90$\,km\,s$^\mathrm{-1}$ and increased to $\sim$160\,km\,s$^\mathrm{-1}$ on Jan 21.

One month after the maximum we obtained two additional SCORPIO spectra in consecutive nights and one OSIRIS spectrum two days later. At that time the V brightness of the star decreased by $\sim$1.5 mag and the colors drastically reddened. Correspondingly, the spectrum of \object{M31N 2015-01a} moved to the lower temperature classes. Now, the neutral elements absorptions dominate the spectrum together with weak TiO molecular bands. Among the strongest absorptions in the spectrum are the \ion{Na}{I} doublet, \ion{Ba}{II} 6142\,\AA\ and 6497\,\AA\, \ion{Ca}{I} 6573\,\AA\, confirmed by \citet{2015arXiv150407747W}, and the near IR triplet of \ion{Ca}{II}. The \ion{Ca}{II} H \& K lines are detected in emission. The strong \ion{Na}{I} lines are well split in our OSIRIS spectrum. This allowed us to estimate their FWHM$\sim200\pm15$\,km\,s$^\mathrm{-1}$ and their radial velocity to $\sim-570\pm15$\,km\,s$^\mathrm{-1}$. In good accordance with the velocity measured in all our February spectra of \object{M31N 2015-01a} (see below) and testifies to the stellar origin of these absorptions. In the OSIRIS spectrum H$\alpha$ presents as a very weak emission divided in two components by a central absorption. In the SCORPIO spectrum only a marginal emission component can be seen. A comparison of the \object{M31N 2015-01a} spectrum about one month after the brightness maximum with the stellar spectra libraries of \citet{1984ApJS...56..257J} and \citet{2003A&A...402..433L} shows closest similarity with a K3-4I spectral type. Cross-correlating the February spectra with a K3I template we estimated an average velocity $-588\pm40$\,km\,s$^\mathrm{-1}$, suggesting an increase of the ejected matter velocity to $\sim$220\,km\,s$^\mathrm{-1}$. The cross-correlation of the SCORPIO spectra with the OSIRIS one shows small velocity differences of $8.4\pm6.6$\,km\,s$^\mathrm{-1}$ and $-10\pm14$\,km\,s$^\mathrm{-1}$ for Feb 21 and 22 respectively. All uncertainties are $1\sigma$ equivalent. 

\begin{figure}
\centering
\includegraphics[width=8.4cm]{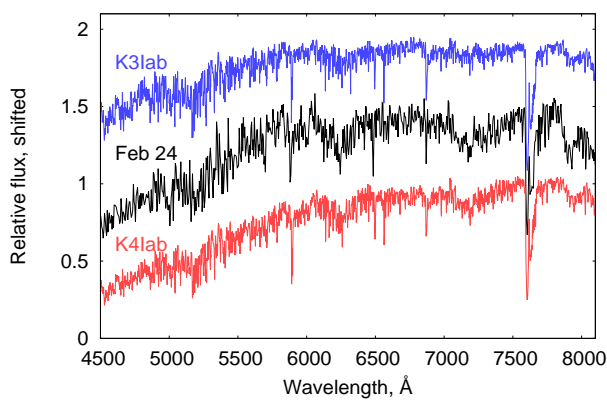}
\caption{The combined GTC/OSIRIS spectrum dereddened by $E_{B-V}=0.4$\,mag in comparison to HD 157999 (K3Iab) and 41 Gem (K4Iab).}
\label{gtcspec} 
\end{figure}

Using the task \textsc{fitspec} in the \textsc{stsdas} \textsc{synphot}\footnote{\textsc{stsdas} is a product of the Space Telescope Science Institute, which is operated by AURA for NASA.} and template spectra from \citet{1984ApJS...56..257J}, \citet{2006MNRAS.371..703S} and \citet{2003A&A...402..433L} we fitted our spectra in an attempt to estimate the reddening. Fitting the spectrum on Jan 15 with a F5I template we obtained a reddening $E_\mathrm{B-V}=0.37$\,mag. For the spectrum on Jan 21 fitted with a F0I template the corresponding  $E_\mathrm{B-V}$ was 0.33\,mag. Merged spectra on Feb 21 and 22 and the spectrum on Feb 24 were best fitted with a K3I template and $E_\mathrm{B-V}=0.25$\,mag and $E_\mathrm{B-V}=0.42$\,mag respectively.

Combining the column hydrogen densities of \citet{2006A&A...453..459N} with the gas-to-dust \citep{2011AIPC.1356...45N} ratio yields total color excess on the line of sight $E_{B-V}=0.44\pm0.08$\,mag. An identical result: $E_{B-V}=0.42\pm0.03$\,mag is derived from the maps of the M31 dust surface density \citep{2014ApJ...780..172D}. A comparison of the $B-V$ and $V-R$ colors with the catalog compiled by \citet{2011ApJS..193....1W} shows that the reddening $E_{B-V}$ is in the range of $0.15-0.58$\,mag. The mean reddening value, obtained from the spectra -- $E_{B-V}=0.35\pm0.10$\,mag, of which $\sim$0.05\,mag can be attributed to the interstellar reddening in our Galaxy \citep{2011ApJ...737..103S}, is well within this range. It suggests a true color of $(B-V)_{0}=0.15\pm0.10$\,mag and a $V$-band bolometric correction of +0.03\,mag at maximum. Adopting a distance modulus $(m-M)=24.47$\,mag\footnote{The NED database (\url{http://ned.ipac.caltech.edu/}) lists a median $(m-M)=24.45$\,mag\ for M31. If the object lies in the M31 plane, its location suggests a +0.02 mag difference from the center.} and a total-to-selective extinction $R_V=3.1$ yields a peak absolute magnitude $M_V=-10.13\pm0.30$\,mag. This sets the luminosity limits at maximum on $L_{bol}=8.7^{+3.3}_{-2.2}\times10^{5}L_{\odot}$. Note, that M31N 2015-01a is projected inside a tiny (2$\arcsec$x2$\arcsec$) HII region \#1527 \citep{2011AJ....142..139A}. The latter list an extinction $A_{R}=0.538$\,mag, which corresponds to the somewhat lower $E_{B-V}=0.23$\,mag. An association with this HII region could decrease the luminosity estimate to $\sim6\times10^{5}L_{\odot}$.

\section{Summary}

 The outburst of M31N 2015-01a was a very luminous event, reaching $\sim10^{6}L_{\odot}$ at the optical maximum. We present accurate photometry and astrometry in the period January-March 2015. The transient shares all major observational characteristics of red novae and we conclude that it belongs to this group of events. The fading was slow at first with $t_{2}\sim50\,d$ in $R$-band. In a one-month period $B-R$ increased by 2.1 mag and the observed spectral type changed from F0I to K3I. The exploration of red novae is so far based on a very small sample of sources, which makes M31N 2015-01a so important: it moves us a step closer to recognizing similarities and differences in their behaviour. 

\begin{acknowledgements}
      This work was partially financed by grant No. BG051 PO001-3.3.06-0057 of the European Social Fund. Based on observations made with the GTC telescope, in the Spanish Observatorio del Roque de los Muchachos of the Instituto de Astrof\'{i}sica de Canarias, under Director's Discretionary Time. Also based on observations made with the Nordic Optical Telescope at the Observatorio del Roque de los Muchachos. This work makes use of observations from the LCOGT network. KH was supported by the project RVO:67985815. AK\&EO gratefully acknowledge observing grant support from the Institute of Astronomy and Rozhen National Astronomical Observatory, Bulgarian Academy of Sciences. KV acknowledges support from the Hungarian Research Grant OTKA K-109276, OTKA K-113117 and the "Lend\"ulet" Program (LP2012-31) of the Hungarian Academy of Sciences. KV, KS and LM have been supported by the Lend\"ulet-2009 program of the Hungarian Academy of Sciences and the ESA PECS Contract No. 4000110889/14/NL/NDe. EAB and VPG thank the Russian Foundation for Basic Research for the financial support by Grant 14-02-00759. The research was supported by the Russian Scientific Foundation (grant N\,14-50-00043). SF acknowledges support of the Russian Government Program of Competitive Growth of Kazan Federal University. AWS acknowledges support from NSF grant AST1009566. We thank to H. Ku\v{c}\'akov\'a, J. Vra\v{s}til and M. Wolf for obtaining images at Ond\v{r}ejov.
\end{acknowledgements}


\bibliographystyle{aa}
\bibliography{m31_rednova}

\begin{thebibliography}{39}
\expandafter\ifx\csname natexlab\endcsname\relax\def\natexlab#1{#1}\fi

\bibitem[{{Afanasiev} \& {Moiseev}(2005)}]{2005AstL...31..194A}
{Afanasiev}, V.~L. \& {Moiseev}, A.~V. 2005, Astronomy Letters, 31, 194

\bibitem[{{Azimlu} {et~al.}(2011){Azimlu}, {Marciniak}, \&
  {Barmby}}]{2011AJ....142..139A}
{Azimlu}, M., {Marciniak}, R., \& {Barmby}, P. 2011, \aj, 142, 139

\bibitem[{{Bond} {et~al.}(2003){Bond}, {Henden}, {Levay}, {Panagia}, {Sparks},
  {Starrfield}, {Wagner}, {Corradi}, \& {Munari}}]{2003Natur.422..405B}
{Bond}, H.~E., {Henden}, A., {Levay}, Z.~G., {et~al.} 2003, \nat, 422, 405

\bibitem[{{Cepa}(1998)}]{1998Ap&SS.263..369C}
{Cepa}, J. 1998, \apss, 263, 369

\bibitem[{{Dong} {et~al.}(2015){Dong}, {Kochanek}, {Adams}, \&
  {Prieto}}]{2015ATel.7173....1D}
{Dong}, S., {Kochanek}, C.~S., {Adams}, S., \& {Prieto}, J.-L. 2015, ATel,
  7173, 1

\bibitem[{{Draine} {et~al.}(2014){Draine}, {Aniano}, {Krause}, {Groves},
  {Sandstrom}, {Braun}, {Leroy}, {Klaas}, {Linz}, {Rix}, {Schinnerer},
  {Schmiedeke}, \& {Walter}}]{2014ApJ...780..172D}
{Draine}, B.~T., {Aniano}, G., {Krause}, O., {et~al.} 2014, \apj, 780, 172

\bibitem[{{Evans} {et~al.}(2003){Evans}, {Geballe}, {Rushton}, {Smalley}, {van
  Loon}, {Eyres}, \& {Tyne}}]{2003MNRAS.343.1054E}
{Evans}, A., {Geballe}, T.~R., {Rushton}, M.~T., {et~al.} 2003, \mnras, 343,
  1054

\bibitem[{{Fabrika} {et~al.}(2015){Fabrika}, {Barsukova}, {Valeev},
  {Vinokurov}, {Sholukhova}, {Goranskij}, {Hornoch}, {Henze}, \&
  {Shafter}}]{2015ATel.6985....1F}
{Fabrika}, S., {Barsukova}, E.~A., {Valeev}, A.~F., {et~al.} 2015, ATel, 6985,
  1

\bibitem[{{Goranskij} {et~al.}(2015){Goranskij}, {Cherjasov}, {Safonov},
  {Vosyakova}, {Barsukova}, {Spiridonova}, \& {Valeev}}]{2015ATel.7206....1G}
{Goranskij}, V.~P., {Cherjasov}, D.~V., {Safonov}, B.~S., {et~al.} 2015, ATel,
  7206, 1

\bibitem[{{Hamuy} {et~al.}(1994){Hamuy}, {Suntzeff}, {Heathcote}, {Walker},
  {Gigoux}, \& {Phillips}}]{1994PASP..106..566H}
{Hamuy}, M., {Suntzeff}, N.~B., {Heathcote}, S.~R., {et~al.} 1994, \pasp, 106,
  566

\bibitem[{{Hamuy} {et~al.}(1992){Hamuy}, {Walker}, {Suntzeff}, {Gigoux},
  {Heathcote}, \& {Phillips}}]{1992PASP..104..533H}
{Hamuy}, M., {Walker}, A.~R., {Suntzeff}, N.~B., {et~al.} 1992, \pasp, 104, 533

\bibitem[{{Hodgkin} {et~al.}(2015){Hodgkin}, {Campbell}, {Fraser}, {Jonker},
  {Torres}, {Wevers}, {van Velzen}, {Rixon}, {Koposov}, {Walton},
  {Wyrzykowski}, {Kostrzewa-Rutkowska}, {Baltay}, {Ellman}, {Rabinowitz},
  {Rostami}, {Feindt}, {Kowalski}, \& {Nugent}}]{2015ATel.6952....1H}
{Hodgkin}, S.~T., {Campbell}, H., {Fraser}, M., {et~al.} 2015, ATel, 6952, 1

\bibitem[{{Jacoby} {et~al.}(1984){Jacoby}, {Hunter}, \&
  {Christian}}]{1984ApJS...56..257J}
{Jacoby}, G.~H., {Hunter}, D.~A., \& {Christian}, C.~A. 1984, \apjs, 56, 257

\bibitem[{{Jockers} {et~al.}(2000){Jockers}, {Credner}, {Bonev}, {Kisele},
  {Korsun}, {Kulyk}, {Rosenbush}, {Andrienko}, {Karpov}, {Sergeev}, \&
  {Tarady}}]{2000KFNTS...3...13J}
{Jockers}, K., {Credner}, T., {Bonev}, T., {et~al.} 2000, Kinematika i Fizika
  Nebesnykh Tel Supplement, 3, 13

\bibitem[{{Kaminski} {et~al.}(2015){Kaminski}, {Menten}, {Tylenda}, {Hajduk},
  {Patel}, \& {Kraus}}]{2015arXiv150306570K}
{Kaminski}, T., {Menten}, K.~M., {Tylenda}, R., {et~al.} 2015, ArXiv e-prints
  [\eprint[arXiv]{1503.06570}]

\bibitem[{{Kochanek} {et~al.}(2014){Kochanek}, {Adams}, \&
  {Belczynski}}]{2014MNRAS.443.1319K}
{Kochanek}, C.~S., {Adams}, S.~M., \& {Belczynski}, K. 2014, \mnras, 443, 1319

\bibitem[{{Kulkarni} {et~al.}(2007){Kulkarni}, {Ofek}, {Rau}, {Cenko},
  {Soderberg}, {Fox}, {Gal-Yam}, {Capak}, {Moon}, {Li}, {Filippenko}, {Egami},
  {Kartaltepe}, \& {Sanders}}]{2007Natur.447..458K}
{Kulkarni}, S.~R., {Ofek}, E.~O., {Rau}, A., {et~al.} 2007, \nat, 447, 458

\bibitem[{{Kurtenkov} {et~al.}(2015{\natexlab{a}}){Kurtenkov}, {Ovcharov},
  {Nedialkov}, {Kostov}, {Bachev}, {Dimitrova}, {Popov}, \&
  {Valcheva}}]{2015ATel.6941....1K}
{Kurtenkov}, A., {Ovcharov}, E., {Nedialkov}, P., {et~al.} 2015{\natexlab{a}},
  ATel, 6941, 1

\bibitem[{{Kurtenkov} {et~al.}(2015{\natexlab{b}}){Kurtenkov}, {Tomov},
  {Fabrika}, {Barsukova}, {Valeev}, {Pessev}, {Vida}, {Molnar}, {Sarneczky},
  {Goranskij}, {Hornoch}, {Henze}, {Shafter}, {Ovcharov}, {Nedialkov},
  {Kostov}, {Valenti}, \& {Stritzinger}}]{2015ATel.7150....1K}
{Kurtenkov}, A., {Tomov}, T., {Fabrika}, S., {et~al.} 2015{\natexlab{b}}, ATel,
  7150, 1

\bibitem[{{Le Borgne} {et~al.}(2003){Le Borgne}, {Bruzual}, {Pell{\'o}},
  {Lan{\c c}on}, {Rocca-Volmerange}, {Sanahuja}, {Schaerer}, {Soubiran}, \&
  {V{\'{\i}}lchez-G{\'o}mez}}]{2003A&A...402..433L}
{Le Borgne}, J.-F., {Bruzual}, G., {Pell{\'o}}, R., {et~al.} 2003, \aap, 402,
  433

\bibitem[{{Magnier} {et~al.}(1992){Magnier}, {Lewin}, {van Paradijs},
  {Hasinger}, {Jain}, {Pietsch}, \& {Truemper}}]{1992A&AS...96..379M}
{Magnier}, E.~A., {Lewin}, W.~H.~G., {van Paradijs}, J., {et~al.} 1992, \aaps,
  96, 379

\bibitem[{{Martini} {et~al.}(1999){Martini}, {Wagner}, {Tomaney}, {Rich},
  {della Valle}, \& {Hauschildt}}]{1999AJ....118.1034M}
{Martini}, P., {Wagner}, R.~M., {Tomaney}, A., {et~al.} 1999, \aj, 118, 1034

\bibitem[{{Mason} {et~al.}(2010){Mason}, {Diaz}, {Williams}, {Preston}, \&
  {Bensby}}]{2010A&A...516A.108M}
{Mason}, E., {Diaz}, M., {Williams}, R.~E., {Preston}, G., \& {Bensby}, T.
  2010, \aap, 516, A108

\bibitem[{{Massey} {et~al.}(2006){Massey}, {Olsen}, {Hodge}, {Strong},
  {Jacoby}, {Schlingman}, \& {Smith}}]{2006AJ....131.2478M}
{Massey}, P., {Olsen}, K.~A.~G., {Hodge}, P.~W., {et~al.} 2006, \aj, 131, 2478

\bibitem[{{Munari} {et~al.}(2002){Munari}, {Henden}, {Kiyota}, {Laney},
  {Marang}, {Zwitter}, {Corradi}, {Desidera}, {Marrese}, {Giro}, {Boschi}, \&
  {Schwartz}}]{2002A&A...389L..51M}
{Munari}, U., {Henden}, A., {Kiyota}, S., {et~al.} 2002, \aap, 389, L51

\bibitem[{{Nedialkov} {et~al.}(2011){Nedialkov}, {Williams}, {Green}, \&
  {Hatzidimitriou}}]{2011AIPC.1356...45N}
{Nedialkov}, P., {Williams}, B., {Green}, J., \& {Hatzidimitriou}, D. 2011, in
  AIP Conference Series, Vol. 1356, AIP Conference Series, ed. I.~{Zhelyazkov}
  \& T.~{Mishonov}, 45--49

\bibitem[{{Nieten} {et~al.}(2006){Nieten}, {Neininger}, {Gu{\'e}lin},
  {Ungerechts}, {Lucas}, {Berkhuijsen}, {Beck}, \&
  {Wielebinski}}]{2006A&A...453..459N}
{Nieten}, C., {Neininger}, N., {Gu{\'e}lin}, M., {et~al.} 2006, \aap, 453, 459

\bibitem[{{Oke}(1990)}]{1990AJ.....99.1621O}
{Oke}, J.~B. 1990, \aj, 99, 1621

\bibitem[{{Rau} {et~al.}(2007){Rau}, {Kulkarni}, {Ofek}, \&
  {Yan}}]{2007ApJ...659.1536R}
{Rau}, A., {Kulkarni}, S.~R., {Ofek}, E.~O., \& {Yan}, L. 2007, \apj, 659, 1536

\bibitem[{{Rich} {et~al.}(1989){Rich}, {Mould}, {Picard}, {Frogel}, \&
  {Davies}}]{1989ApJ...341L..51R}
{Rich}, R.~M., {Mould}, J., {Picard}, A., {Frogel}, J.~A., \& {Davies}, R.
  1989, \apjl, 341, L51

\bibitem[{{Roeser} {et~al.}(2010){Roeser}, {Demleitner}, \&
  {Schilbach}}]{2010AJ....139.2440R}
{Roeser}, S., {Demleitner}, M., \& {Schilbach}, E. 2010, \aj, 139, 2440

\bibitem[{{S{\'a}nchez-Bl{\'a}zquez} {et~al.}(2006){S{\'a}nchez-Bl{\'a}zquez},
  {Peletier}, {Jim{\'e}nez-Vicente}, {Cardiel}, {Cenarro},
  {Falc{\'o}n-Barroso}, {Gorgas}, {Selam}, \& {Vazdekis}}]{2006MNRAS.371..703S}
{S{\'a}nchez-Bl{\'a}zquez}, P., {Peletier}, R.~F., {Jim{\'e}nez-Vicente}, J.,
  {et~al.} 2006, \mnras, 371, 703

\bibitem[{{Schlafly} \& {Finkbeiner}(2011)}]{2011ApJ...737..103S}
{Schlafly}, E.~F. \& {Finkbeiner}, D.~P. 2011, \apj, 737, 103

\bibitem[{{Shumkov} {et~al.}(2015){Shumkov}, {Lipunov}, {Gorbovskoy},
  {Tiurina}, {Balanutsa}, {Kuznetsov}, {Pruzhinskaya}, {Denisenko}, {Rufanov},
  {Vladimirov}, {Ivanov}, {Yazev}, {Budnev}, {Gress}, {Poleshchuk},
  {Parkhomenko}, {Tlatov}, {Dormidontov}, {Yurkov}, {Sergienko}, {Varda},
  {Sinyakov}, {Gabovich}, {Krushinsky}, {Zalozhnih}, {Popov}, {Bourdanov},
  {Buckley}, {Potter}, {Kniazev}, {Kotze}, {Shumkov}, {Vladimirov}, {Lipunov},
  {Gorbovskoy}, {Tiurina}, {Balanutsa}, {Kuznetsov}, {Pruzhinskaya},
  {Denisenko}, {Rufanov}, {Ivanov}, {Yazev}, {Budnev}, {Gress}, {Poleshchuk},
  {Parkhomenko}, {Tlatov}, {Dormidontov}, {Yurkov}, {Sergienko}, {Varda},
  {Sinyakov}, {Gabovich}, {Krushinsky}, {Zalozhnih}, {Popov}, {Bourdanov},
  {Buckley}, {Potter}, {Kniazev}, {Kotze}, \&
  {Shurpakov}}]{2015ATel.6911....1S}
{Shumkov}, V., {Lipunov}, V., {Gorbovskoy}, E., {et~al.} 2015, ATel, 6911, 1

\bibitem[{{Sparks} {et~al.}(2008){Sparks}, {Bond}, {Cracraft}, {Levay},
  {Crause}, {Dopita}, {Henden}, {Munari}, {Panagia}, {Starrfield}, {Sugerman},
  {Wagner}, \& {White}}]{2008AJ....135..605S}
{Sparks}, W.~B., {Bond}, H.~E., {Cracraft}, M., {et~al.} 2008, \aj, 135, 605

\bibitem[{{Tylenda} {et~al.}(2011){Tylenda}, {Hajduk}, {Kami{\'n}ski},
  {Udalski}, {Soszy{\'n}ski}, {Szyma{\'n}ski}, {Kubiak}, {Pietrzy{\'n}ski},
  {Poleski}, {Wyrzykowski}, \& {Ulaczyk}}]{2011A&A...528A.114T}
{Tylenda}, R., {Hajduk}, M., {Kami{\'n}ski}, T., {et~al.} 2011, \aap, 528, A114

\bibitem[{{Tylenda} \& {Soker}(2006)}]{2006A&A...451..223T}
{Tylenda}, R. \& {Soker}, N. 2006, \aap, 451, 223

\bibitem[{{Williams} {et~al.}(2015){Williams}, {Darnley}, {Bode}, \&
  {Steele}}]{2015arXiv150407747W}
{Williams}, S.~C., {Darnley}, M.~J., {Bode}, M.~F., \& {Steele}, I.~A. 2015,
  ArXiv e-prints [\eprint[arXiv]{1504.07747}]

\bibitem[{{Worthey} \& {Lee}(2011)}]{2011ApJS..193....1W}
{Worthey}, G. \& {Lee}, H.-c. 2011, \apjs, 193, 1

\end{thebibliography}

\onltab{
\begin{table}
\begin{center}
\caption[Calculated magnitudes for M31N 2015-01a in Johnson-Cousins $BVRI$ filters.]{Calculated magnitudes for M31N 2015-01a in Johnson-Cousins $BVRI$ filters.}\label{table_phot}
\small
\begin{tabular}{c@{}c@{ }c@{ }c@{ }c@{ }c@{ }c@{ }c@{ }c@{}}
\hline\hline
 JD-2457000 \, & filter \, & mag \, &  err  \, & telescope \, & airmass \\
\hline
    36.264 \, & B \, &  19.641 \, & 0.161 \, & RSCH    \, &  1.17  \\
    38.248 \, & B \, &  17.046 \, & 0.017 \, & BTA     \, &  1.30  \\
    38.345 \, & B \, &  16.992 \, & 0.029 \, & RSCH    \, &  1.69  \\
    39.261 \, & B \, &  16.549 \, & 0.017 \, & RSCH    \, &  1.19  \\
    39.413 \, & B \, &  16.498 \, & 0.029 \, & UL50    \, &  1.53  \\
    40.295 \, & B \, &  16.215 \, & 0.017 \, & RSCH    \, &  1.35  \\
    44.196 \, & B \, &  15.934 \, & 0.017 \, & BTA     \, &  1.16  \\
    46.612 \, & B \, &  16.109 \, & 0.026 \, & LCOGT   \, &  1.32  \\
    48.658 \, & B \, &  16.679 \, & 0.242 \, & LCOGT   \, &  1.71  \\
    51.611 \, & B \, &  17.261 \, & 0.026 \, & LCOGT   \, &  1.38  \\
    60.323 \, & B \, &  17.865 \, & 0.096 \, & KSCH    \, &  1.79  \\
    61.252 \, & B \, &  17.974 \, & 0.026 \, & KSCH    \, &  1.32  \\
    67.257 \, & B \, &  18.294 \, & 0.066 \, & KSCH    \, &  1.42  \\
    68.246 \, & B \, &  18.383 \, & 0.022 \, & RSCH    \, &  1.53  \\
    72.249 \, & B \, &  18.719 \, & 0.103 \, & KSCH    \, &  1.46  \\
    75.224 \, & B \, &  18.791 \, & 0.018 \, & BTA     \, &  1.92  \\
    93.340 \, & B \, &  20.076 \, & 0.051 \, & NOT     \, &  2.53  \\
    96.346 \, & B \, &  20.385 \, & 0.041 \, & NOT     \, &  2.99  \\
\hline
    38.246 \, & V \, &  16.416 \, & 0.008 \, & BTA     \, &  1.29  \\
    44.187 \, & V \, &  15.436 \, & 0.011 \, & BTA     \, &  1.14  \\
    66.319 \, & V \, &  16.626 \, & 0.012 \, & KSCH    \, &  1.93  \\
    67.243 \, & V \, &  16.621 \, & 0.026 \, & KSCH    \, &  1.35  \\
    72.241 \, & V \, &  16.832 \, & 0.065 \, & KSCH    \, &  1.41  \\
    75.225 \, & V \, &  16.873 \, & 0.013 \, & BTA     \, &  1.93  \\
    89.323 \, & V \, &  17.869 \, & 0.059 \, & UL50    \, &  1.92  \\
    92.246 \, & V \, &  18.098 \, & 0.108 \, & KSCH    \, &  1.91  \\
\hline
    36.243 \, & R \, &  18.087 \, & 0.026 \, & RSCH    \, &  1.11  \\
    37.210 \, & R \, &  17.213 \, & 0.011 \, & R2M     \, &  1.05  \\
    38.246 \, & R \, &  16.051 \, & 0.011 \, & BTA     \, &  1.29  \\
    38.330 \, & R \, &  16.047 \, & 0.009 \, & RSCH    \, &  1.55  \\
    39.242 \, & R \, &  15.679 \, & 0.009 \, & RSCH    \, &  1.13  \\
    40.283 \, & R \, &  15.407 \, & 0.009 \, & RSCH    \, &  1.29  \\
    40.390 \, & R \, &  15.364 \, & 0.012 \, & R2M     \, &  2.38  \\
    44.189 \, & R \, &  15.133 \, & 0.011 \, & BTA     \, &  1.14  \\
    46.617 \, & R \, &  15.165 \, & 0.013 \, & LCOGT   \, &  1.34  \\
    48.661 \, & R \, &  15.347 \, & 0.032 \, & LCOGT   \, &  1.75  \\
    51.615 \, & R \, &  15.609 \, & 0.011 \, & LCOGT   \, &  1.41  \\
    57.285 \, & R \, &  15.678 \, & 0.011 \, & ONDR    \, &  1.36  \\
    59.238 \, & R \, &  15.692 \, & 0.011 \, & ONDR    \, &  1.20  \\
    60.317 \, & R \, &  15.675 \, & 0.012 \, & KSCH    \, &  1.73  \\
    61.234 \, & R \, &  15.683 \, & 0.011 \, & KSCH    \, &  1.24  \\
    62.596 \, & R \, &  15.681 \, & 0.009 \, & LCOGT   \, &  1.50  \\
    64.608 \, & R \, &  15.689 \, & 0.009 \, & LCOGT   \, &  1.66  \\
    66.310 \, & R \, &  15.725 \, & 0.011 \, & KSCH    \, &  1.82  \\
    67.239 \, & R \, &  15.741 \, & 0.009 \, & KSCH    \, &  1.33  \\
    68.230 \, & R \, &  15.804 \, & 0.013 \, & RSCH    \, &  1.42  \\
    72.244 \, & R \, &  15.839 \, & 0.027 \, & KSCH    \, &  1.42  \\
    73.250 \, & R \, &  15.858 \, & 0.009 \, & ONDR    \, &  1.40  \\
    75.223 \, & R \, &  15.894 \, & 0.024 \, & BTA     \, &  1.90  \\
    80.270 \, & R \, &  16.111 \, & 0.011 \, & ONDR    \, &  1.68  \\
    86.329 \, & R \, &  16.417 \, & 0.021 \, & ONDR    \, &  2.72  \\
    87.302 \, & R \, &  16.437 \, & 0.016 \, & ONDR    \, &  2.28  \\
    89.266 \, & R \, &  16.642 \, & 0.012 \, & ONDR    \, &  1.88  \\
    89.337 \, & R \, &  16.617 \, & 0.061 \, & UL50    \, &  2.08  \\
    92.245 \, & R \, &  16.987 \, & 0.104 \, & KSCH    \, &  1.90  \\
    93.345 \, & R \, &  17.011 \, & 0.011 \, & NOT     \, &  2.68  \\
    96.351 \, & R \, &  17.315 \, & 0.013 \, & NOT     \, &  3.19  \\
    99.283 \, & R \, &  17.617 \, & 0.035 \, & ONDR    \, &  2.54  \\
   105.279 \, & R \, &  18.499 \, & 0.084 \, & ONDR    \, &  2.79  \\
\hline
    60.320 \, & I \, &  15.028 \, & 0.021 \, & KSCH    \, &  1.76  \\
    61.245 \, & I \, &  15.017 \, & 0.018 \, & KSCH    \, &  1.28  \\
    67.241 \, & I \, &  15.019 \, & 0.018 \, & KSCH    \, &  1.34  \\
    68.272 \, & I \, &  15.085 \, & 0.016 \, & RSCH    \, &  1.78  \\
    72.246 \, & I \, &  15.048 \, & 0.028 \, & KSCH    \, &  1.44  \\
    89.357 \, & I \, &  15.344 \, & 0.111 \, & UL50    \, &  2.34  \\
    92.248 \, & I \, &  15.377 \, & 0.065 \, & KSCH    \, &  1.92  \\
    93.350 \, & I \, &  15.557 \, & 0.021 \, & NOT     \, &  2.84  \\
    96.356 \, & I \, &  15.735 \, & 0.021 \, & NOT     \, &  3.42  \\
\hline
\end{tabular} 
\end{center}  
\end{table}   
}

\end{document}